\documentclass[12pt]{article}
\usepackage{graphics} 
\usepackage{cite}
\newcommand   {\etal}    {{\it et~al.}}
\makeatletter \renewcommand\@biblabel[1]{} \makeatother

\newcommand{\beq}{\begin{equation}}
\newcommand{\eeq}{\end{equation}}
\newcommand{\beqa}{\begin{eqnarray}}
\newcommand{\eeqa}{\end{eqnarray}}
\newcommand{\bea}{\begin{eqnarray}}
\newcommand{\eea}{\end{eqnarray}}
\newcommand   {\ev}[1]   {\langle #1\rangle}
\newcommand   {\bl}      {\sigma^{(s)}_k}
\newcommand   {\fl}      {\psi^{(s)}}

\newcommand   {\hir}     {h_i^{(r)}}

\newcommand   {\dcr}     {\Delta\chi^{(r)}}

\input{psfig}

\begin{document}

\begin{flushright}
LU TP 99-40\\
Revised version\\
September 19, 2000
\end{flushright}

\vspace{0.4in}

\LARGE
\begin{center}
{\bf On Hydrophobicity Correlations\\in Protein Chains\\} 
\vspace{0.3in}
\large
Anders Irb\"ack\footnote{irback@thep.lu.se} and 
Erik Sandelin\footnote{erik@thep.lu.se}\\   
\vspace{0.10in}
Complex Systems Division, Department of Theoretical Physics\\ 
University of Lund,  S\"olvegatan 14A,  S-223 62 Lund, Sweden \\
{\tt http://www.thep.lu.se/tf2/complex/}\\
\vspace{0.3in}	

Submitted to {\it Biophysical Journal}
\end{center}

\vspace{0.6in}

\normalsize
Abstract:\\
We study the statistical properties of hydrophobic/polar 
model sequences with unique native states on the square
lattice. It is shown that this ensemble of sequences differs 
from random sequences in significant ways in terms of both the 
distribution of hydrophobicity along the chains and  
total hydrophobicity. Whenever statistically 
feasible, the analogous calculations are performed for
a set of real enzymes, too.\\


\newpage

\section{Introduction}

Functional protein sequences exhibit the ability to fold spontaneously 
into a unique native state (Creighton, 1993). A natural
step in order to understand this crucial property is to 
compare good and bad folding sequences in simple models 
where conformational space can be properly explored. 
Most such studies have been directed toward identifying physical 
characteristics of good folders, and in this important area some 
progress has been made (S\u{a}li \etal, 1994; Bryngelson \etal, 
1995; Klimov and Thirumalai, 1998; Nymeyer \etal, 1998).  
In this paper we address the question of how good folders 
differ from random sequences in purely statistical terms. A related 
but different topic is how sequences that share the same (unique) 
native state are distributed in sequence space. This question and  
its evolutionary implications have recently attracted considerable 
attention (Li \etal, 1996; Bornberg-Bauer, 1997; 
Govindarajan and Goldstein, 1997a,b; 
Bastolla \etal, 1999; Broglia \etal; 1999; 
Bornberg-Bauer and Chan, 1999; Tiana \etal, 2000). 

In a recent study of a hydrophobic/polar off-lattice model, it was 
found that good folders tend to show negative hydrophobicity
correlations along the chains (Irb\"ack \etal, 1997). The 
analogous calculations gave, morevover, qualitatively similar 
results for a major class of real proteins, 
corresponding to typical total hydrophobicities
(Irb\"ack \etal, 1996). On the other hand, the opposite behavior, 
positive hydrophobicity correlations, has been reported for a class
of designed model sequences that display certain protein-like features
(Khokhlov and Khalatur, 1998, 1999). These designed sequences are, 
for instance, not meant to have unique native states, so the different 
results do not represent a contradiction. However, it shows that 
sequence correlations in proteins is a delicate issue that requires 
a careful analysis.     

The main goal of this paper is to test the robustness of the conclusion 
that good folding model sequences as well as functional proteins 
show negative hydrophobicity correlations. To this end we perform new 
calculations for both model and real sequences. The model we study 
is the minimal HP model on the square lattice (Lau and Dill, 1989; 
Dill \etal, 1995). This choice makes it possible for us to improve
significantly on the statistics in the previous study (Irb\"ack \etal, 1997), 
which was based on an off-lattice model. The real sequences studied 
are single-domain enzymes taken from the CATH protein structure 
classification database (Orengo \etal, 1997),  which we hope displays 
statistical properties representative of functional (globular) 
folding units. With this restriction on protein type, it turns out that 
the previous, somewhat artificial, restriction on total hydrophobicity 
(Irb\"ack \etal, 1996) can be lifted.    

\section{Methods}\label{sec:2}

\subsection{Sequences}\label{sec:2.1}

Let us first define the sequences studied. The real sequences 
studied are the 173 nonhomologous single domain enzymes found 
in the October 1998 release of the CATH database (Orengo \etal, 1997). 
These sequences are transformed into binary hydrophobicity strings,
by taking the six amino acids Leu, Ile, Val, Phe, Met, and Trp as 
hydrophobic ($\sigma_i=1$) and the others as hydrophilic ($\sigma_i=-1$).
This choice is somewhat arbitrary. Therefore, we also tried a 20-valued 
hydrophobicity scale, which did not affect any of the conclusions
below. In CATH, the most general level of classification is denoted ``class'' 
and describes the relative content of $\alpha$ helices and $\beta$ sheets. 
Below, the class dependence of our results is checked by separate 
calculations for each of the three major classes: mainly $\alpha$, 
mainly $\beta$, and $\alpha\beta$. A fourth class, low secondary 
structure content, exists but it is not considered separately, 
as only 3 of the 173 sequences belong to it. In our calculations
we also divide the sequences into extracellular and intracellular ones. 
Following Martin \etal~(1998), we take the presence of a disulphide 
bridge as an indicator of extracellular location. The number of enzymes 
in the different subsets studied can be found in Table~\ref{tab:3} below.

The model we use is the minimal two-dimensional HP model 
(Lau and Dill, 1989), whose behavior 
is known in quite some detail (Dill \etal, 1995). It contains
only two types of amino acids, H (hydrophobic, $\sigma_i=1$) and 
P (polar, $\sigma_i=-1$), and the chain conformation is represented  
as a self-avoiding walk on a lattice. The formation of a 
hydrophobic core is favored by defining the energy as  
minus the number of HH pairs that are nearest neighbors 
on the lattice but not along the chain. On the square lattice, 
it turns out that this simple choice of energy function 
is sufficient in order to get a significant number of sequences 
with unique ground states (Chan and Dill, 1994; Irb\"ack and Sandelin, 1998); 
complete enumeration of all possible sequences and structures shows that
the fraction of such sequences is roughly 2\% for $N\le 18$.
Throughout this paper we consider all HP sequences that have
unique ground states as good folding sequences. Also 
central is that the sequences are able to 
fold fast into their native states,  
a requirement that we ignore. This is a reasonable simplification 
because the sequences are short and because almost all have the same 
energy gap between ground state and next lowest level.     

\subsection{Sequence Correlations}\label{sec:2.2} 

Our statistical analysis of hydrophobicity strings  
can be divided into two parts. The first part deals with
the distribution of hydrophobicity along the chains; how
does a ``good'' sequence with length $N$ and total
hydrophobicity 
\beq
M=\sum_{i=1}^N\sigma_i
\label{M}\eeq
differ from a typical sequence with the same $N$ and $M$?
This question can be addressed by monitoring  
variables such as the number of hydrophobic and hydrophilic 
clumps along the chain (White and Jacobs, 1990), Fourier 
amplitudes (Irb\"ack \etal, 1996), or random walk (Brownian bridge)
representations (Pande \etal, 1994). In this paper we work with    
block variables, a widely used technique that has proven useful  
in studies of DNA sequences (Peng \etal, 1992) as well as proteins 
(Irb\"ack \etal, 1996).            

In addition to the distribution of hydrophobicity along the 
chains, we also study the distribution of the total hydrophobicity $M$. 
This analysis relies entirely on comparisons between 
observed sequences, which makes it statistically more difficult, 
especially for the real sequences with varying $N$.

\subsubsection{The Blocking Method}\label{sec:2.2.1} 

In this method, for a given size $s$, the sequence is divided 
into blocks each consisting of $s$ consecutive $\sigma_i$ along 
the chain. The block variable $\bl$ is then defined as the sum of 
the $s$ $\sigma_i$ values in block $k$ ($k=1,\ldots,N/s$). A useful      
quantity is the mean-square fluctuation    
\beq
\fl={s\over N}
\sum_{k=1}^{N/s}\fl_k \qquad\qquad 
\fl_k={1\over K}\left(\bl-sM/N\right)^2\qquad\qquad   
\label{msblock}\eeq
where we choose the normalization factor
\beq
K=\frac{N^2-M^2}{N^2-N}(1-s/N)\,.
\label{K}\eeq  
With this choice, the average of $\fl$ over all possible
sequences with given $N$ and $M$ takes the simple form (Irb\"ack \etal, 1996)
\ \beq
\ev{\fl}_{N,M}=s\,,
\label{random}\eeq 
independent of $N$ and $M$. 

\subsubsection{The Distribution of Total Hydrophobicity}\label{sec:2.2.2}

We study the $M$ distribution for different fixed $N$, focusing   
on the mean $\ev{M}_N$ (the subscript indicates fixed $N$) and the 
normalized variance
\beq
\chi=\frac{1}{N}\left\langle\left(M-\ev{M}_N\right)^2\right\rangle_N\,.
\label{chi}\eeq
It is easily verified that  
\beq
\chi=\frac{4}{N}\sum_{i=1}^Nh_i(1-h_i)+\frac{1}{N}\sum_{i\ne j}c_{ij}\,,
\label{diag}\eeq
where $h_i=(1+\ev{\sigma_i}_N)/2$ denotes
the fraction of sequences that have $\sigma_i=1$, and 
$c_{ij}=\ev{\sigma_i\sigma_j}_N-\ev{\sigma_i}_N\ev{\sigma_j}_N$
is the $\sigma_i,\sigma_j$ correlation. So, if the $\sigma_i$ values are 
uncorrelated, then 
\beq
\chi=\chi_1\equiv\frac{4}{N}\sum_{i=1}^Nh_i(1-h_i)\,,
\label{chi1}\eeq
which becomes  
\beq
\chi=\chi_0\equiv4h(1-h)
\label{chi0}\eeq
in case the hydrophobicity profile $\{h_i\}$ is flat with 
$h_i=h$ for all $i$. Below these two predictions are tested  
for the model sequences. 

Unfortunately, our set of enzymes cannot be analyzed this way, 
due to limited statistics. However, as we will see, it turns 
out that the data for the mean $\ev{M}_N$ can be approximately 
described by a simple linear relation, 
$\ev{M}_N\approx \bar{M}=(2\bar{h}-1)N$. As an effective measure 
of the fluctuations in $M$, we therefore consider   
\beq
\bar\chi=\left\langle\left(
\frac{M-\bar{M}}{N^{1/2}}\right)^2\right\rangle\,,  
\label{chia}\eeq
where the average now is over all sequences, irrespective of $N$. 
If the $\sigma_i$ values, for each $N$, were uncorrelated with
identical $h_i=\bar{h}$, then we would have
\beq
\bar\chi=\bar\chi_0\equiv4\bar{h}(1-\bar{h})\,.
\label{chia0}\eeq   

Let us finally stress that $\fl$ and $\chi$ are 
fundamentally different measurements. In the blocking   
method individual sequences are compared to random
sequences with the same $N$ and $M$. Hence, $\fl$ 
provides direct information on the distribution of 
$\sigma_i=\pm 1$ {\it along} the chains. This is not
true for $\chi$ and the correlation $c_{ij}$. This correlation is 
not necessarily physical. The behavior of the     
analogue of $c_{ij}$ in the ordered phase of an Ising 
magnet provides an illustration of this. In this case, 
$c_{ij}$ does not vanish at large distance, although 
the physical correlation length is finite.    

\subsection{Individual Structures}\label{sec:2.3}

As mentioned in the introduction, several recent model studies 
have addressed the question of how sequences that fold to 
the same native state are related. In particular, 
using an HP-like model with compact structures
only, Li~\etal~(1996) found that structure-preserving
mutations tend to be largely independent for highly designable 
structures. To see whether this behavior is consistent 
with our analysis, we perform two measurements for different 
fixed structures, too.

Consider a given structure $r$, and let $\{h_i^{(r)}\}$ be the corresponding 
hydrophobicity profile ($h_i^{(r)}$ is the probability that 
$\sigma_i=1$). The first quantity we calculate is   
\beq
\dcr=\chi^{(r)}-\frac{4}{N}\sum_{i=1}^N\hir(1-\hir)\,,
\label{deltachi}\eeq 
where $\chi^{(r)}$ is defined as $\chi$ in Eq.~\ref{chi} 
but for fixed structure. $\dcr$  measures the average 
$\sigma_i,\sigma_j$ correlation for fixed structure
(see Eq.~\ref{diag}).  The second quantity is the 
entropy
\beq
S=-\sum_{i=1}^N \left[\hir \ln\hir + \left(1-\hir\right)
\ln\left(1-\hir\right)\right]
\eeq
for a system of independent $\sigma_i$ with hydrophobicity
profile $\{h_i^{(r)}\}$. If the $\sigma_i$ values are approximately 
independent,  then ${\rm e}^S$ provides an order-of-magnitude estimate 
of the actual number of sequences, $N_r$. If this is not the case, 
then ${\rm e}^S$ overestimates $N_r$. 

\section{Results}\label{sec:3}

In this section we present the results of our analyses of 
the mean-square block fluctuations $\fl$ and the distribution 
of total hydrophobicity, $M$, for model and real sequences. 
We end the section with some comments on our model results 
and related studies of similar models.  

\subsection{The Blocking Method}

\subsubsection{Model Sequences}  

\begin{figure}
\begin{center}
\vspace{-7mm}
\hspace{35mm}

\rotatebox{270}{\psfig{figure=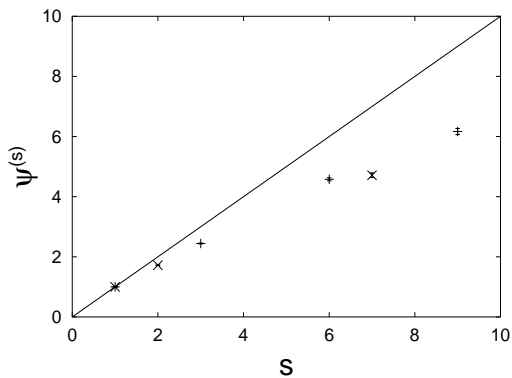,width=5.0cm,height=7.0cm}}

\caption{The mean-square block fluctuation $\fl$ against
block size $s$ for good $N=18$ sequences in the HP model.
Shown are results both for the full sequences $(+)$ and for 
the subsequences consisting of the central 14 amino acids 
($\times$). The straight line represents random 
sequences; see Eq.~\protect\ref{random}.} 
\label{fig:1}
\end{center}	
\end{figure}

In our block variable analysis of HP sequences, we consider 
the 6349 $N=18$ sequences that have unique native states,
which can be obtained by exhaustive enumeration 
(Chan and Dill, 1994). The results 
are compared to expected values for random sequences, 
as described in Sec.~\ref{sec:2.2.1}. This comparison makes sense 
only if the hydrophobicity profile $\{h_i\}$ is uniform. 
From Table~\ref{tab:1} it can be seen that $h_i$ is 
approximately constant in the mid part but  
increases towards the ends. As a check, we therefore calculate 
the mean-square block fluctuation $\fl$ 
in two ways for each sequence: first, for the full sequence; 
and second, after elimination of two amino acids at each end. 
Figure~\ref{fig:1} shows the results of both these calculations. 
We see that the average $\fl$ is smaller than for random sequences, 
irrespective of whether the endpoints are included 
or not. The conclusion that $\fl$, on average, 
is suppressed for good sequences is in perfect agreement 
with earlier results for a different model 
(Irb\"ack \etal, 1996, 1997).

\begin{table}
\begin{center}
\begin{tabular}{ccccccccc}
$h_1$ & $h_2$ & $h_3$ & $h_4$ & $h_5$ & $h_6$ & $h_7$ & $h_8$ & $h_9$\\
\hline
0.794 & 0.642 & 0.467 & 0.456 & 0.553 & 0.498 & 0.526 & 0.479 & 0.523  
\end{tabular}
\caption{Hydrophobicity profile $\{h_i\}$ for good $N=18$ sequences 
in the HP model. By symmetry, $h_i=h_{19-i}$.}
\label{tab:1}
\end{center}
\end{table}

\subsubsection{Enzymes}

\begin{figure}
\begin{center}
\vspace{-7mm}
\hspace{35mm}

\rotatebox{270}{\psfig{figure=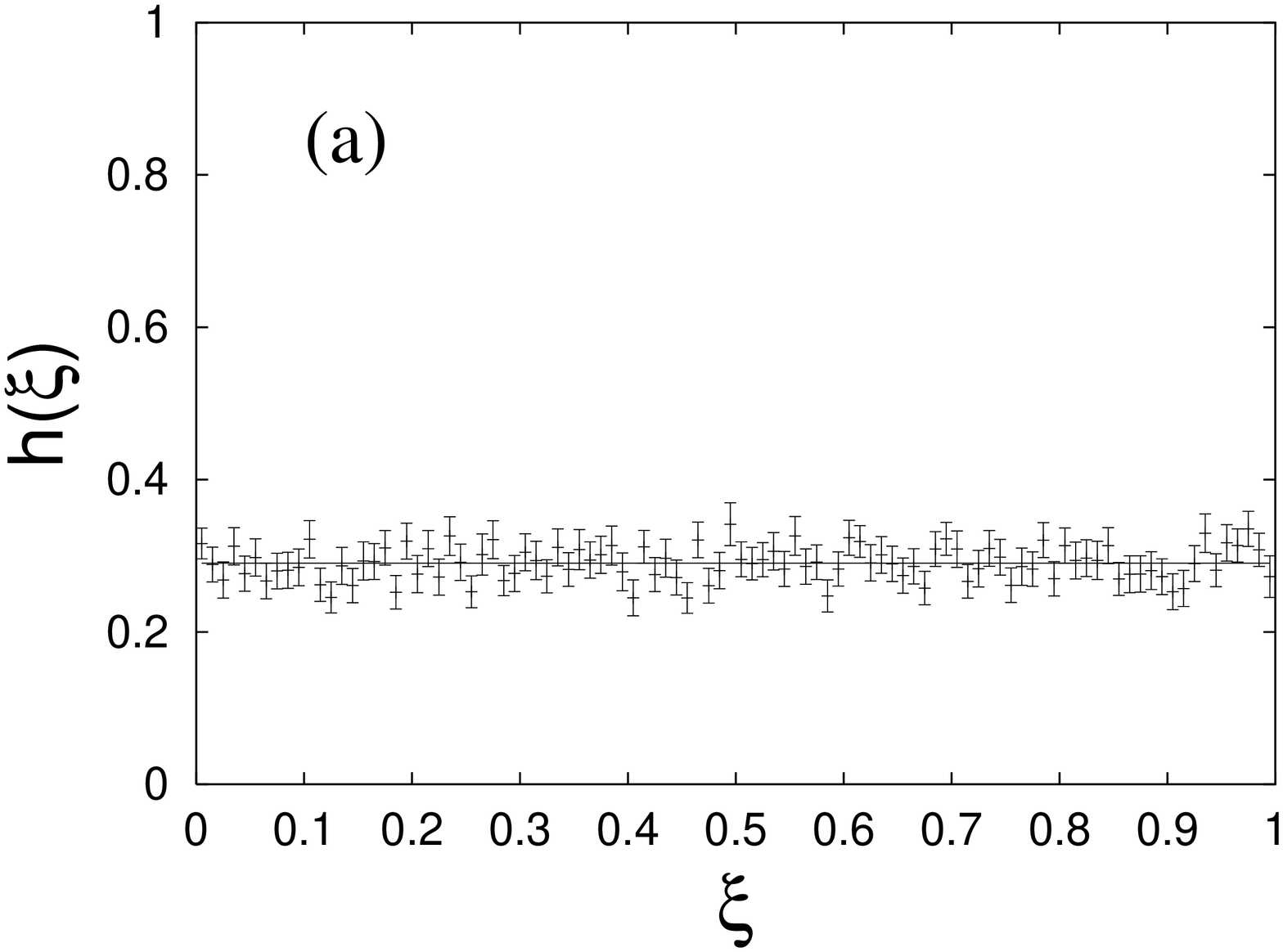,width=5cm,height=7cm}}
\rotatebox{270}{\psfig{figure=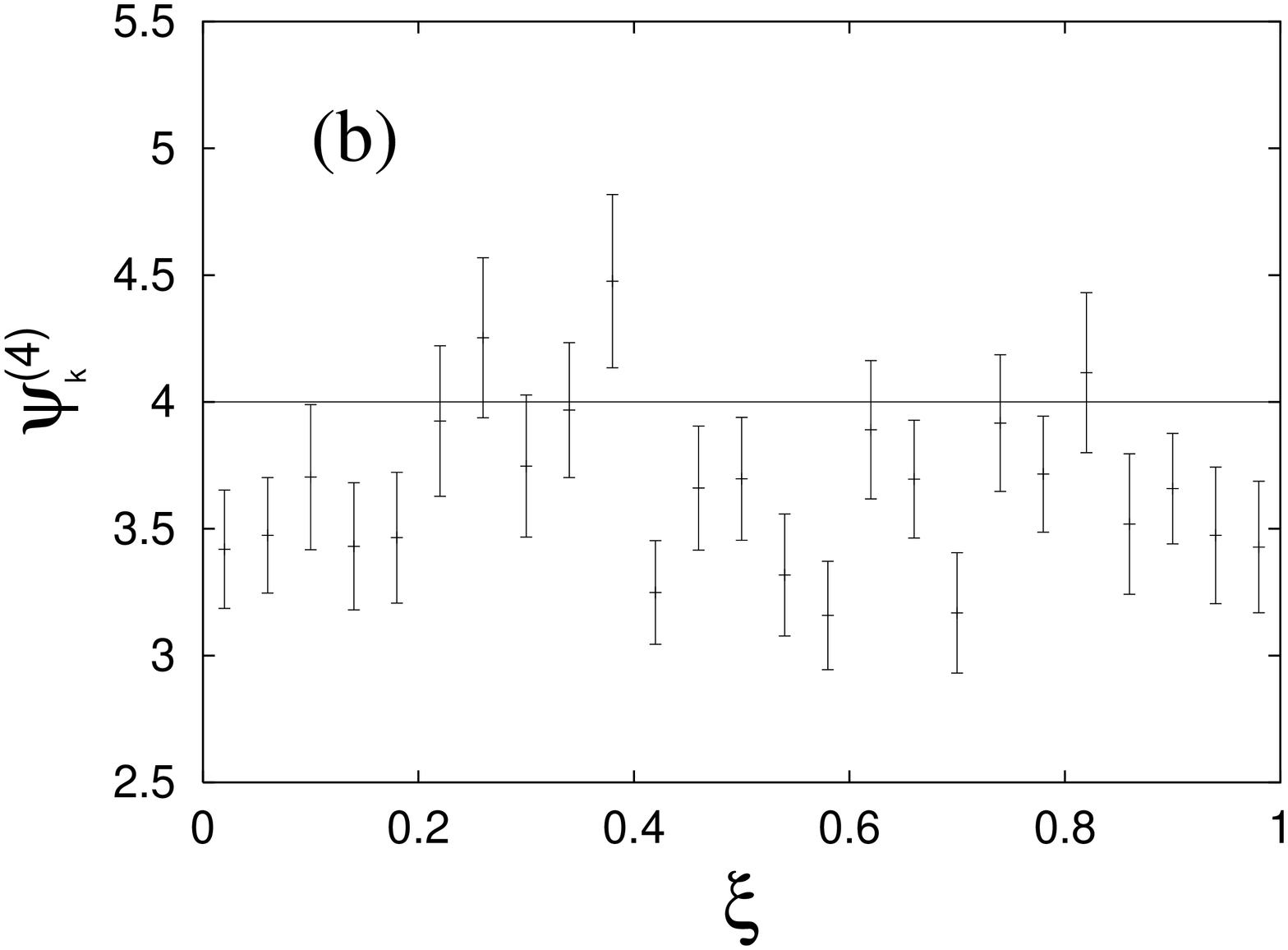,width=5cm,height=7cm}}
\caption{(a) Hydrophobicity profile $h(\xi)$ for the enzymes. 
The horizontal line indicates the mean $\bar{h}\approx0.29$. (b) 
$\psi_k^{(4)}$ as a function of $\xi$ for the enzymes. 
The horizontal line represents random sequences.} 
\label{fig:2}
\end{center}
\end{figure}

We now repeat essentially the same analysis for the enzymes.
The only difference is that, because $N$ is not fixed,  
the hydrophobicity profile $h(\xi)$ is taken to be a function 
of the relative position $\xi$ along the chains. To calculate 
$h(\xi)$, we divide the interval in $\xi$ from 0 (N end) to 1 
(C end) into 100 bins. The results obtained are shown in 
Fig.~\ref{fig:2}a. We see that $h(\xi)$ is approximately
constant throughout the interval $0\le\xi\le1$. 

In an earlier block analysis of functional protein sequences 
(Irb\"ack \etal, 1996), in which there was no restriction on
protein type, the ends were found to display a different 
behavior than the rest of the sequences, and therefore 
they were removed from the analysis. To check if this 
is true for the present data set, we calculate the average 
of $\psi^{(4)}_k$ (see Eq.~\ref{msblock}) as a function 
of $\xi$, using 25 bins in $\xi$. The results are shown in 
Fig.~\ref{fig:2}b. Although the uncertainties are somewhat 
large, there is no sign of the ends behaving differently.

Given these two findings, we calculate the block fluctuations 
using the full sequences, without any elimination of 
amino acids at the ends.          

\begin{figure}

\vspace{-7mm} 

\mbox{
\rotatebox{270}{\psfig{figure=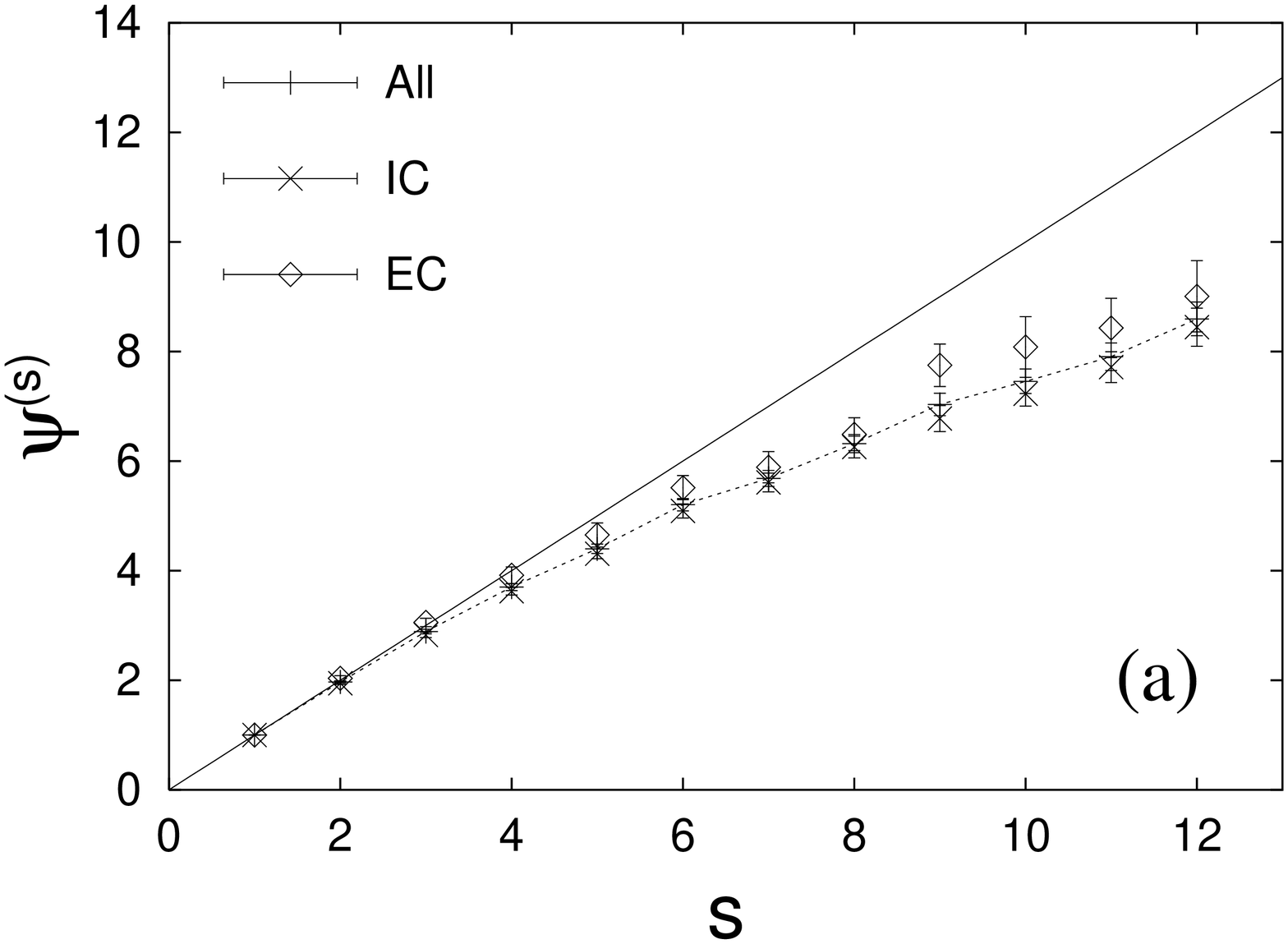,width=5.0cm,height=7.0cm}}
\rotatebox{270}{\psfig{figure=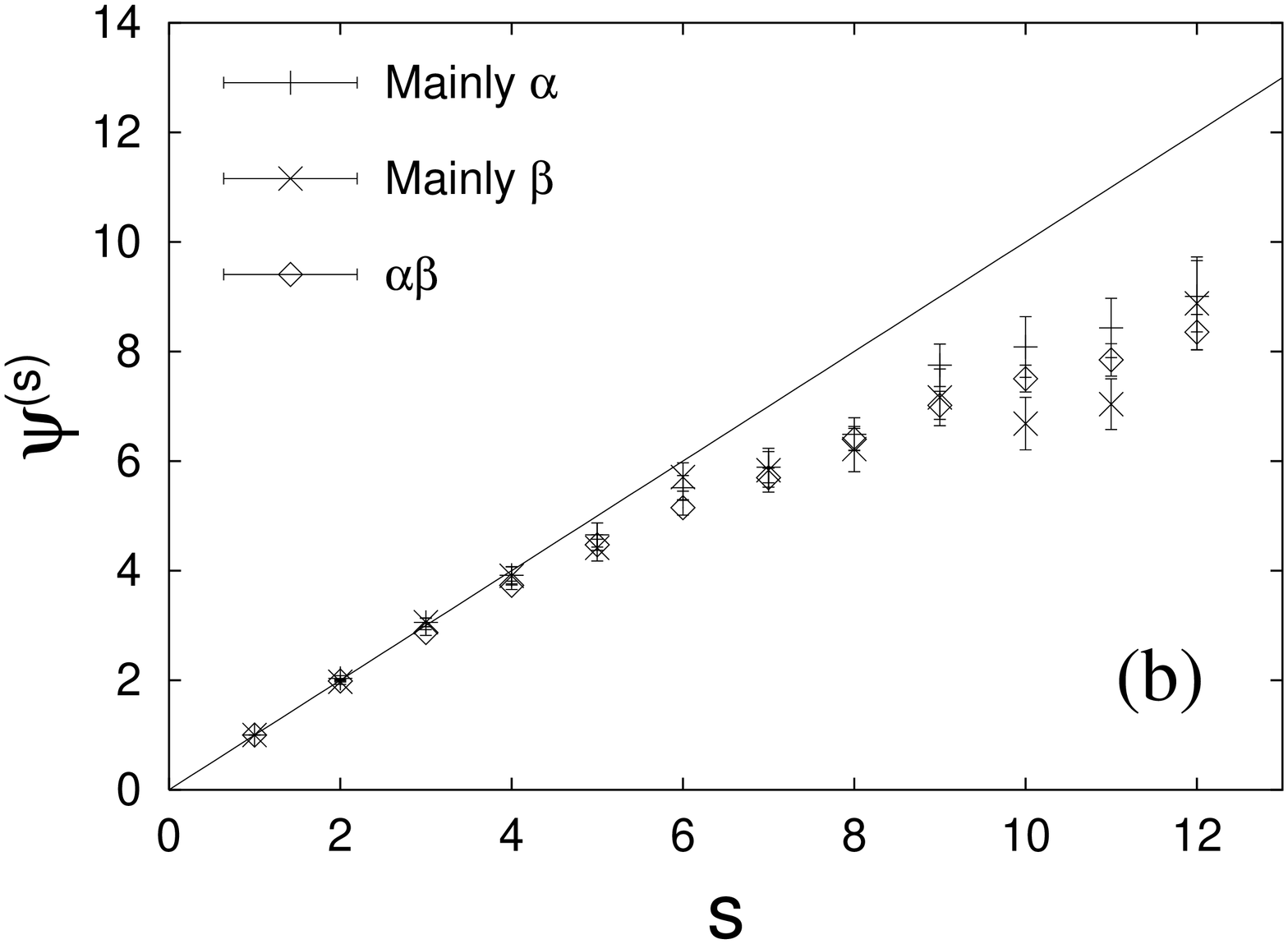,width=5.0cm,height=7.0cm}}
}

\caption{The mean-square block fluctuation $\fl$ against block size $s$
for different groups of enzymes.  
(a) All sequences (data points connected by dashed line) and 
intracellular (IC)/extracellular (EC) sequences. (b) Division 
of the sequences into
three structural classes: mainly $\alpha$, mainly $\beta$, and $\alpha\beta$.
The straight lines represent random sequences.}
\label{fig:3}
\end{figure}

In Fig.~\ref{fig:3} we show the average $\fl$ against block size $s$ 
for the 173 enzymes. Also shown are the results obtained for five 
different subsets of these sequences (see Sec.~\ref{sec:2.1}). We 
see that the results are similar in the different cases, and that  
$\fl$ is smaller than for random sequences. Qualitatively, the          
behavior is similar to that found for the model sequences.  

In this analysis we have chosen to focus on $\fl$. 
Similar deviations from randomness are expected in other 
quantities such as the number of hydrophobic/hydrophilic 
clumps along the chain. The number of clumps tends to 
be large when $\fl$ is small (Irb\"ack \etal, 1997). 

\subsection{The Distribution of Total Hydrophobicity}\label{sec:3.2}

\subsubsection{Model Sequences}  

We now turn to the distribution of the total hydrophobicity $M$.
Table~\ref{tab:2} shows $h=(1+\ev{M}_N/N)/2$ and the normalized 
variance $\chi$ (see Eq.~\ref{chi}) for good HP sequences 
for $N=12,\ldots,18$. Also shown in this table are the two 
predictions $\chi_0$ and $\chi_1$ defined in 
Sec.~\ref{sec:2.2.2}, and a prediction $\chi_2$ that will be 
explained below. Note that $h$ depends quite weakly on $N$. This
implies that the 
fraction of hydrophobic amino acids, unlike the core to surface
ratio of compact chains, does not increase with $N$. 
Of course, it would be interesting to see 
whether this trend persists for much larger $N$.

\begin{table}
\begin{center}
\begin{tabular}{cccccc}
$N$ & $h$ & $\chi$ & $\chi_0$ &  $\chi_1$ & $\chi_2$ \\
\hline
 12  & 0.527 & 0.577  & 0.997 & 0.913 & 0.589 \\
 13  & 0.507 & 0.550  & 1.000 & 0.937 & 0.553 \\
 14  & 0.519 & 0.684  & 0.999 & 0.924 & 0.688 \\
 15  & 0.556 & 0.594  & 0.987 & 0.959 & 0.593 \\
 16  & 0.542 & 0.687  & 0.993 & 0.936 & 0.663 \\
 17  & 0.555 & 0.695  & 0.988 & 0.961 & 0.639 \\
 18  & 0.548 & 0.718  & 0.991 & 0.949 & 0.646 \\

\end{tabular}
\caption{$h=(1+\ev{M}_N/N)/2$ and the normalized variance $\chi$ of $M$
for good HP sequences for different $N$. Also shown are the three
predictions $\chi_0$ (see Eq.~\protect\ref{chi0}), $\chi_1$
(Eq.~\protect\ref{chi1}) and $\chi_2$ (see Sec.~\protect\ref{sec:3.3}).} 
\label{tab:2}
\end{center}
\end{table}           

From Table~\ref{tab:2} we see that $\chi$ is 
smaller than $\chi_0$, which implies that the $\sigma_i$ values
are not both uncorrelated and uniformly distributed. 
Comparing to $\chi_1$ shows that the major part of  
this difference is due to correlations rather 
than non-uniformity. The fact that $\chi<\chi_1$ means 
that the average $c_{ij}$ ($i\ne j$) is negative.  

The two measurements $h$ and $\chi$ are, of course, not enough 
to fully characterize the distribution of good sequences.  
To get an idea of how much information they provide, 
we may compare to the one-dimensional Ising distribution
\beq
P(\sigma) \propto \exp\left(K_1\sum_i\sigma_i\sigma_{i+1}+
K_2\sum_i\sigma_i\right)\ .
\eeq
The measured values of $h$ and $\chi$ for good $N=18$ sequences can be  
reproduced by choosing $K_1\approx-0.16$ and $K_2\approx0.13$. 
For these parameters it turns out that ${\rm e}^S\approx 1.9\times10^5$,
$S$ being the entropy, which means that the effective 
number of sequences contained in $P(\sigma)$ is considerably 
larger than the number of good $N=18$ sequences, 6349.
 
\subsubsection{Enzymes}

\begin{figure}[t]

\vspace{-10mm}
\hspace{35mm}
\begin{center}
\rotatebox{270}{\psfig{figure=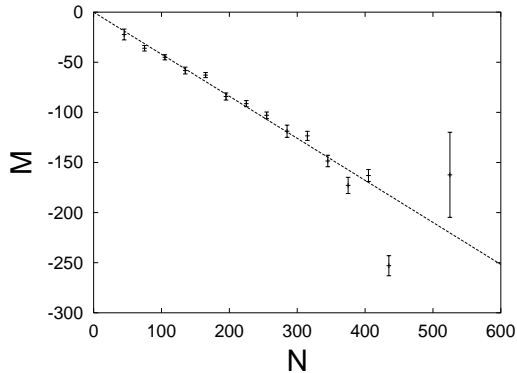,width=5.0cm,height=7.0cm}}

\end{center}
\caption{Total hydrophobicity $M$ against $N$ for the enzymes. 
The data points are averages over intervals of length 30 in $N$. 
The straight line is a least-square fit.}
\label{fig:4}
\end{figure}

To study the $N$ dependence of the total hydrophobicity $M$ 
for the enzymes, we divide the data set into groups corresponding
to different intervals in $N$. Figure~\ref{fig:4} shows the 
average $M$ for these groups against $N$. We see that the 
$N$ dependence is approximately linear. Although the uncertainties 
are difficult to estimate, it is interesting to note that the 
behavior is in perfect agreement with the model results.     

Next we calculate $\bar\chi$ in Eq.~\ref{chia}, using  
$\bar{M}=N(2\bar{h}-1)$ and $\bar{h}=0.29$, as 
obtained from a fit to the data in Fig.~\ref{fig:4}. 
Table~\ref{tab:3} shows $\bar\chi$ for all sequences 
and for the different subgroups described in Sec.~\ref{sec:2.1}. 
We see that $\bar\chi$ for all sequences is larger than 
predicted by Eq.~\ref{chia0}, which contrasts sharply 
with the model results above. We also note that there seems  
to be a strong dependence on group. In particular there appears 
to be a big difference between intra- and extracellular 
enzymes. However, it must be stressed that the 
uncertainties are large. Improved statistics are definitely 
needed in order to draw any firm conclusion about the different 
groups and possible deviations from the model results.

\begin{table}
\begin{center}  

\begin{tabular}{lccc}

Type of chain& No. sequences & $\bar\chi$ & $\bar\chi_0$ \\
\hline
All chains & 173 & 1.50$\pm$0.27 & 0.82\\
Intracellular & 127 & 0.82$\pm$0.13 & 0.83 \\
Extracellular & 46 & 2.92$\pm$1.15 & 0.78 \\
Mainly $\alpha$ & 23 & 1.45$\pm$0.25 & 0.81 \\
Mainly $\beta$ & 39 & 1.63$\pm$0.34 & 0.77 \\
$\alpha\beta$ & 108 & 0.85$\pm$0.14 & 0.83 \\
\end{tabular}
\caption{Analysis of the fluctuations in $M$ for 
the enzymes. The quantities $\bar\chi$ and $\bar\chi_0$ 
are defined by Eqs.~\protect\ref{chia} and 
\protect\ref{chia0}, respectively.}
\label{tab:3}
\end{center}
\end{table}

\subsection{Comments}\label{sec:3.3}

\begin{figure}[t]
\vspace{-41mm}
\mbox{
  \hspace{-35mm}
  \psfig{figure=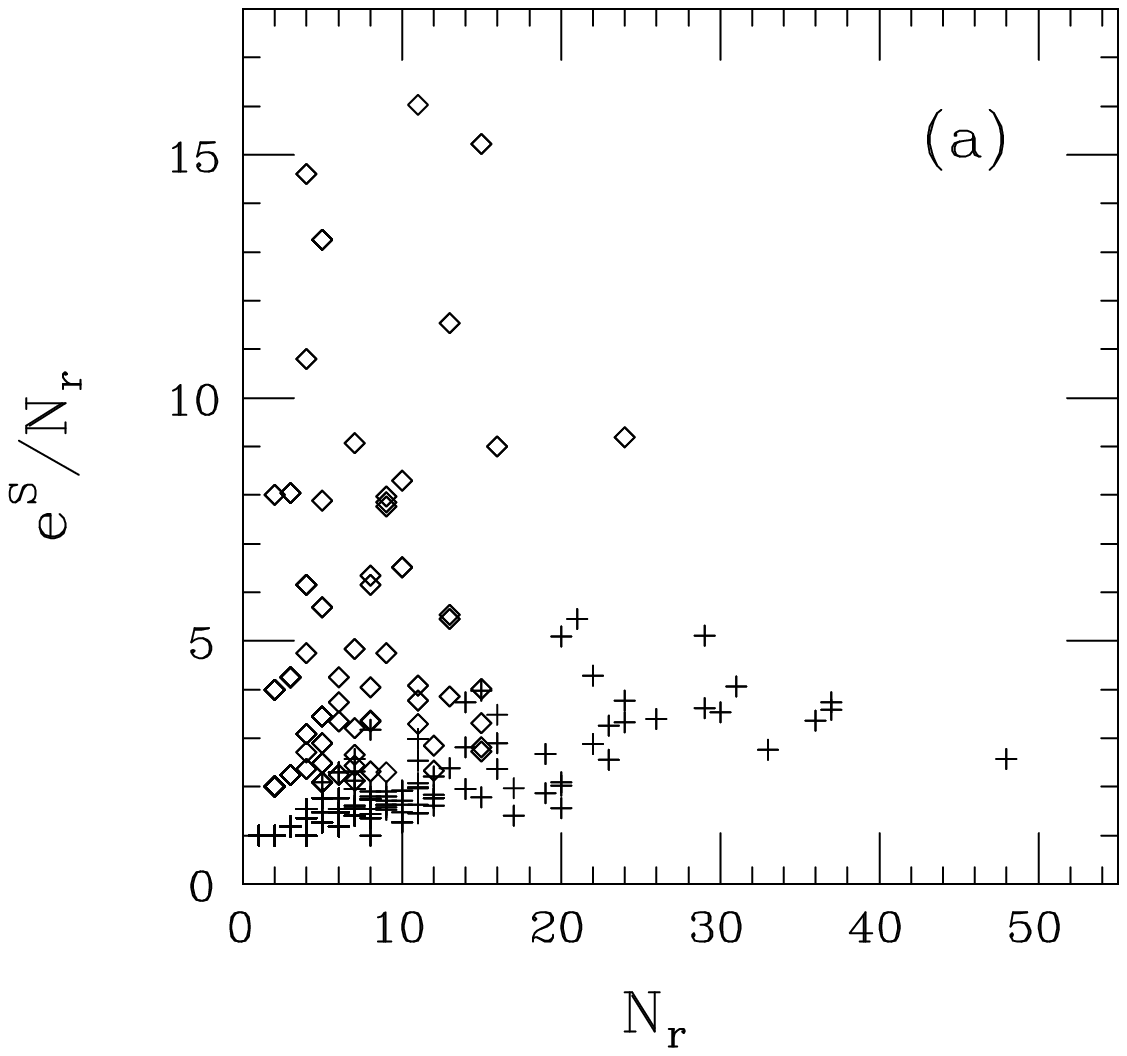,width=10.5cm,height=14cm}
  \hspace{-35mm}
  \psfig{figure=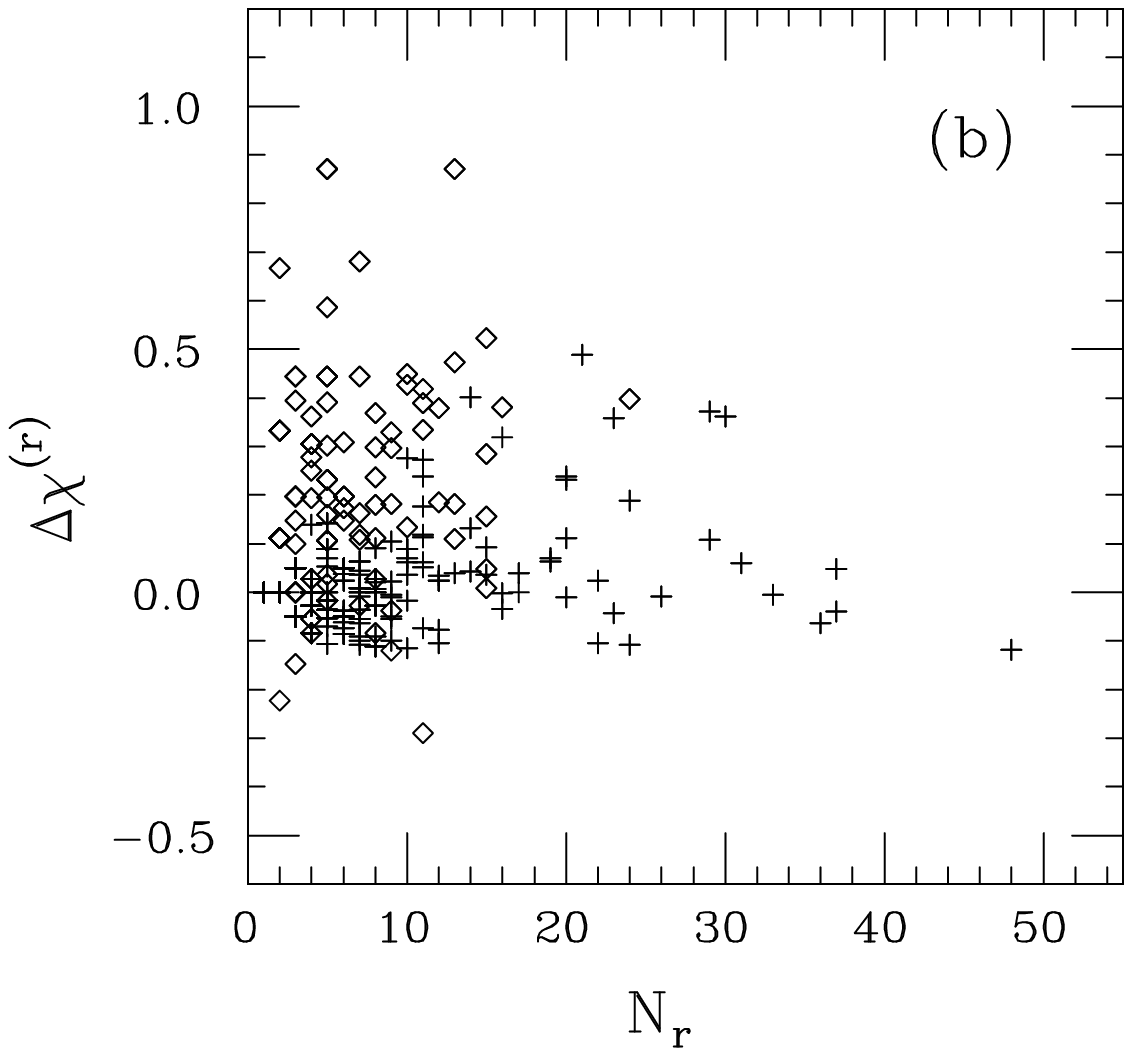,width=10.5cm,height=14cm}
}
\vspace{-41mm}
\caption{(a) ${\rm e}^S/N_r$, $N_r$ and (b) $\dcr$, $N_r$ 
scatter plots for the 1475 designable $N=18$ structures 
in the HP model. The shape of the plot symbol indicates 
whether the sequences form a neutral net ($+$) or 
not ($\diamond$).}
\label{fig:5}
\end{figure}

Our study of HP sequences has been focused on 
structure-independent properties. The question of how 
sequences that share the same (unique) native structure 
are related has recently been examined using similar models 
(Li~\etal, 1996; Bornberg-Bauer, 1997; Bornberg-Bauer and Chan, 1999). 
From these studies, a simple picture seems to emerge for 
structures that are highly designable. For high-$N_r$ structures
($N_r$ is the number of sequences that fold to the structure $r$), 
it has been found that the sequences tend to form a single cluster 
connected by one-point mutations, called a ``neutral net'' 
(Bornberg-Bauer, 1997), and that structure-preserving 
mutations tend to be largely independent (Li~\etal, 1996). 
The latter property was observed in a model with 
compact structures only. We checked that it holds in 
the present model too, which is illustrated in Fig.~\ref{fig:5}. From 
this figure it can be seen that the quantities ${\rm e}^S/N_r$ and 
$|\Delta\chi^{(r)}|$, as defined in Sec.~\ref{sec:2.3}, indeed tend to be 
small for high $N_r$. Also indicated in this figure is whether or not 
the sequences form a neutral net, results first obtained by 
Bornberg-Bauer (1997). 

The fact that structure-preserving mutations are largely 
independent for high $N_r$ does not contradict our previous 
results. To verify this, we calculated $\chi$ from the known 
hydrophobicity profiles 
$\{\hir\}$ under the assumption that the $\sigma_i$ values are independent 
for each structure. The value obtained this way, $\chi_2$, can be found 
in Table~\ref{tab:2} above, and is indeed a relatively good 
approximation to the observed $\chi$.        

Admittedly, the model used in this study is crude. In particular, 
Buchler and Goldstein (1999, 2000) have recently argued,
based on a study of compact lattice chains, that the use 
of a two-letter alphabet leads to designability artifacts, 
which disappear with increasing alphabet size. Let us stress, therefore,
that the analyses discussed in this paper can be 
tested on real proteins in a direct manner. Let us also 
comment on the stability of our results. First, we note that the 
dependence on chain length $N$ is weak. This was explicitly shown 
for $\chi$, and is true for $\fl$ too, although our discussion 
focused on one system size in this case. Second, we note that our 
results are in nice agreement with those obtained earlier using 
a simple hydrophobic/polar off-lattice model (Irb\"ack~\etal, 1997). 
To further explore the model dependence of our results, we 
also did calculations for a ``solvation-like'' two-letter 
model discussed by Ejtehadi~\etal\ (1998a,b) and by Buchler 
and Goldstein (1999, 2000). This model differs from the HP 
model in that the interaction strength is additive 
[$\epsilon($H,H$)=-2\epsilon$, $\epsilon($H,P$)=-\epsilon$ 
and $\epsilon($P,P$)=0$], which means that the total energy 
can be expressed as a simple sum of monomer contributions. 
Buchler and Goldstein argued that HP-like models, unlike 
pair-contact models with larger alphabets, tend to have 
solvation-like designability properties. It is therefore 
interesting to note that when analyzing sequences with 
unique ground states in the solvation-like model defined above, 
we obtained results qualitatively different from those 
for the HP model. More precisely, it turns out that 
the block fluctuations are significantly larger, 
close to random, for the solvation-like model.      

\section{Summary and Discussion}\label{sec:4}

Hydrophobicity plays a key role in the formation of protein 
structures, which makes it of utmost interest to understand 
the statistical distribution of hydrophobicity along the 
chains. In this paper we have analyzed hydrophobic/polar
sequences in the two-dimensional HP lattice model. Whenever 
statistically feasible, the analogous calculations were 
performed for a set of real enzymes, too. Our main findings 
are as follows.    
       
\begin{itemize}
\item Both model sequences and enzymes show mean-square block 
fluctuations $\fl$ that are smaller than for random sequences. 
In particular, this implies that the enzymes display the 
same behavior that had been found previously for general 
proteins with typical total hydrophobicities 
(Irb\"ack \etal, 1996). The present analysis was performed 
without any restriction on total hydrophobicity. 

\item The average total hydrophobicity $M$ varies 
approximately linearly with chain length $N$ over the 
range of $N$ studied, both for model sequences and 
enzymes. This implies, contrary to what one naively might 
expect, that the fraction of hydrophobic amino acids
does not grow with increasing $N$. The fluctuations in $M$ are 
difficult to study for the enzymes, due to statistical uncertainties. 
For the model sequences it turns out that the normalized 
variance $\chi$ is significantly smaller than for random 
sequences.  
\end{itemize} 

We also divided the enzymes into different groups according 
to their structural content, and to whether they reside in an 
intra- or extracellular environment. The fluctuations in total
hydrophobicty appeared to depend on group. However, whether this 
dependence is significant or not is difficult to say, 
due to statistical uncertainties. The mean-square block 
fluctuations are statistcally much easier to measure, and show 
only a weak dependence on group. The conclusion that $\fl$ 
is suppressed is, in particular, the same for all the  
different groups. 
 
A full explanation of the suppression of $\fl$ is probably 
hard to give. Let us note, however, that long hydrophobic 
or hydrophilic stretches in the amino acid sequence 
are likely to lead to degenerate structures, and the suppression 
of sequences containing such stretches should indeed 
tend to make $\fl$ smaller.

The nonrandomness of the block fluctuations provides an indirect 
confirmation of the important role played by hydrophobicity in
the formation of protein structures. Furthermore, it is tempting
to take the similarity with the model results as an indication 
that the ability to form a stable structure represents a significant 
selective advantage in the evolution of proteins.  
It would be interesting to check that the behavior remains the same in  
more realistic models. 

\section*{Acknowledgements}

This work was supported by the Swedish Foundation for Strategic Research.

\newpage

\newcommand  {\Biomet}   {{\it Biometrika}}
\newcommand  {\Biopol}   {{\it Biopolymers}}
\newcommand  {\BC}       {{\it Biophys.\ Chem.\ }}
\newcommand  {\BJ}       {{\it Biophys.\ J.\ }}
\newcommand  {\CPL}      {{\it Chem. Phys. Lett.\ }}
\newcommand  {\COSB}     {{\it Curr.\ Opin.\ Struct.\ Biol.\ }}
\newcommand  {\EL}       {{\it Europhys.\ Lett.\ }}
\newcommand  {\JCC}      {{\it J.\ Comput.\ Chem.\ \ }}
\newcommand  {\JCoP}     {{\it J.\ Comput.\ Phys.\ }}
\newcommand  {\JCP}      {{\it J.\ Chem.\ Phys.\ }}
\newcommand  {\JMB}      {{\it J.\ Mol.\ Biol.\ }}
\newcommand  {\JP}       {{\it J.\ Phys.\ }}
\newcommand  {\JPC}      {{\it J.\ Phys.\ Chem.\ }}
\newcommand  {\JPSJ}     {{\it J. Phys. Soc. (Jap)\ }}
\newcommand  {\JSP}      {{\it J.\ Stat.\ Phys.\ }}
\newcommand  {\JTB}      {{\it J.\ Theor.\ Biol.\ }} 
\newcommand  {\Mac}      {{\it Macromolecules}}
\newcommand  {\MC}       {{\it Makromol.\ Chem.,\ Theory Simul.\ }}
\newcommand  {\MP}       {{\it Molec.\ Phys.\ }}
\newcommand  {\NAR}      {{\it Nucleic\ Acids\ Res.\ }}
\newcommand  {\Nat}      {{\it Nature}}
\newcommand  {\NP}       {{\it Nucl.\ Phys.\ }}
\newcommand  {\Phy}      {{\it Physica\ }}
\newcommand  {\Pro}      {{\it Proteins:\ Struct.\ Funct.\ Genet.\ }}
\newcommand  {\ProEng}   {{\it Protein\ Eng.\ }}
\newcommand  {\Pa}       {{\it Physica}}
\newcommand  {\PL}       {{\it Phys.\ Lett.\ }}
\newcommand  {\PNAS}     {{\it Proc.\ Natl.\ Acad.\ Sci.\ USA}}
\newcommand  {\PR}       {{\it Phys.\ Rev.\ }}
\newcommand  {\PRL}      {{\it Phys.\ Rev.\ Lett.\ }}
\newcommand  {\PRS}      {{\it Proc.\ Roy.\ Soc.\ }}
\newcommand  {\PS}       {{\it Protein\ Sci.\ }}
\newcommand  {\RMP}      {{\it Rev.\ Mod.\ Phys.\ }}
\newcommand  {\Sci}      {{\it Science}}
\newcommand  {\SFD}      {{\it Structure with Folding \& Design}}
\newcommand  {\Str}      {{\it Structure}}
\newcommand  {\ZP}       {{\it Z.\ Physik}}


\end{document}